\documentclass[twocolumn,amsmath,amssymb,prl ]{revtex4}

\usepackage{float}
\usepackage{latexsym} 
\usepackage{amsmath,amsthm}
\usepackage{ifpdf}
\usepackage{epstopdf}
\usepackage{pdfpages}
\usepackage{dcolumn}
\usepackage{bm}
\usepackage{braket}

\usepackage{color}

\newcommand {\Pdert}{\frac {\partial}{\partial t}}

\begin{document}
\def \k{\bold k}
\def \p{\bold p}
\def \r{\bold r}
\def \A{\bold A}
\def \a{\hat a_{\k}}
\def \ad{\hat a^{\dagger}_{\k}}
\def \b{\hat b_{-\k}}
\def \bd{\hat b^{\dagger}_{-\k}}
\def \beq{\begin{equation}}
\def \eeq{\end{equation}}
\def \bea{\begin{eqnarray}}
\def \eea{\end{eqnarray}}
\def \bes{\begin{split}}
\def \ees{\end{split}}
\def \besu{\begin{subequations}}
\def \esu{\end{subequations}}
\def \bea{\begin{align}}
\def \eal{\end{align}}

\title{Coulomb effects in the absorbance spectra of two-dimensional Dirac materials}
\author{Leone Di Mauro Villari}
\author{Ian Galbraith}
\author{Fabio Biancalana}
\affiliation{Institute of Photonics and Quantum Sciences, School of Engineering and Physical Sciences, SUPA, Heriot-Watt University, Edinburgh EH14 4AS, UK.}

\begin{abstract}
A wide range of materials like graphene, topological insulators and transition metal dichalcogenides (TMDs) share an interesting property: the low energy excitations behave as Dirac particles. This emergent behavior of Dirac quasiparticles defines a large class of media that are usually called Dirac materials. The linear and nonlinear optical properties of Dirac materials with a gap are still largely unexplored, and in this Letter we build the foundations of a novel way to study the linear optical properties of these two-dimensional media. Our approach is based on a new Dirac-like formulation of the standard semiconductor Bloch equations used in semiconductor physics. We provide an explicit expression of the linear absorbance -- which we call the relativistic Elliott formula -- and use this to quantify the variation of the continuum absorbance spectrum with the strength of the Coulomb interaction (the Sommerfeld factor). Our calculations also show how the Coulomb enhancements scales with the bandgap and vanishes for zero bandgap, shedding new light on the behaviour of graphene for low light intensities. The results presented are in good quantitative agreement with published experimental results. Our new theory will allow researchers to explore the nonlinear interactions of intense, ultrashort pulses with TMDs, and the framework is flexible enough to be adapted to different experimental situations, such as cavities, multilayers, heterostructures and microresonators.
\end{abstract}
\maketitle

Condensed matter physics is witnessing a rapid expansion in the fabrication of a wide variety of materials with Dirac fermion quasi-particle excitations. These seemingly diverse materials possess properties that are direct consequence of the Dirac spectrum of the quasiparticles and are universal. For example neutral superfluids $d$-wave superconductors and graphene, which are characterised by a massless Dirac fermion low energy spectrum,  all exhibit the same power-law temperature dependence of the fermionic specific heat with the only differences arising from the dimensionality of the excitation phase space  \cite{wb}.
Amongst Dirac materials, gapless graphene is one of the most widely studied due to its unusual physical properties arising from the interplay of  its reduced dimensionality and the nature of its excitation spectrum  \cite{ng,mam,nb,wk}. 

Perhaps the most intriguing  characteristic of graphene is the so-called universal absorbance. The opacity of suspended graphene is defined solely by the fine structure constant ($\alpha=e^2/\hbar c$), the parameter that describes coupling between light and relativistic electrons which is traditionally associated with quantum electrodynamics (QED) rather than materials science  \cite{nb}. This universal behavior is known to be broken by opening a gap, with the appearance of bound states (excitons) due to the electron-hole Coulomb interactions. 
A gap can be opened in graphene when the sample is deposited on a dielectric substrate \cite{zg} or it can be induced by impurities, lattice defects \cite{kb} and mechanical strain \cite{ny}. 
In terms of the low excitation Dirac Hamiltonian the effect of a gap opening is analogous to the chiral symmetry breaking in QED that generates a mass gap in the particle-anti-particle spectrum \cite{cd,mk}. This suggests that a gapped Dirac material can be accurately described by a massive Dirac Hamiltonian. An example of two-dimensional materials whose low energy excitations behave as massive Dirac particles is provided by monolayer transition metal dichalcogenides (TMDs) such as molybdenum disulfide (MoS$_2$) and tungsten diselenide (WSe$_2$) \cite{sk}. Akin to graphene, these materials display new physical properties, distinct from their bulk counterparts \cite{ml,cr,xl,ub,lc}.  Here, emission is dominated by excitons and trions due to the strong Coulomb interactions arising from their low dimensionality and reduced dielectric screening. Remarkably, despite their tiny atomic thickness they can absorb up to 20\% of the incident light \cite{lc,ub}. The proper analysis of the measured spectra requires the identification of the optically active states. For this purpose, one derives dipole allowed selection rules that result from the conservation of the total angular momentum in the excitation and emission processes. In conventional direct gap GaAs-type quantum wells the radial nature of the dipole moment imposes that only $s$-type excitonic states couple to light \cite{hk,sk}. This leads to the well-known excitonic Rydberg series. It has been shown  that in the case of monolayer TMDs  in the regime of strong Coulomb interactions the system collapses into a excitonic insulator phase and optically bright $p$-excitons are allowed \cite{sk}. However recent works pointed out that for a gapped Dirac system the symmetry properties of the $\bold K$-point result in a non radial dipole moment with a nontrivial angular dependence related to the appearance of a Berry phase \cite{cb}. This peculiar characteristic suggests that different dipole allowed transitions should be present and optically active $p$-states should appear even in the weak Coulomb interaction regime. 

 In this Letter we calculate the absorbance spectra of Dirac materials in the relatively new theoretical framework of the instantaneous eigenstates \cite{I} and we apply this formalism to predict the spectra of realistic materials. We derive the Elliott formula for a gapped Dirac system taking into account both the discrete and the continuum part of the Wannier eigenvalue problem. In particular we study how the absorbance spectrum evolves as the bandgap is reduced to zero. We show that even in the presence of relatively strong electron-hole Coulomb interaction the gap reduction leads to the universal absorbance of graphene and we show how this behavior is closely related to the structure of the relativistic hydrogenic problem. 
 
 Before detailing our calculations we show in figure \ref{fig1} the  absorbance spectrum calculated for the Molibdenum disulfide (MoS$_2$). We  observe a high absorbance value ($15$\%) around the energy gap  in agreement with previous experimental and theoretical  results \cite{lc,xv,r}.  
Further agreement with experimental data can be observed in the strength of the $1s$ exciton peak ($5$\%) and the value of the binding energy \cite{ub}.  
The intensity of the continuum absorbance peak is around three time stronger than the $1s$ exciton one. This is a peculiar characteristic shared by Dirac materials and in contrast to non-Dirac materials.  In what follows we show how this spectrum relates to the relativistic structure of the low-energy quasiparticle excitations.  
 \begin{figure}[h] 
\begin{center}
\includegraphics[width=1 \columnwidth]{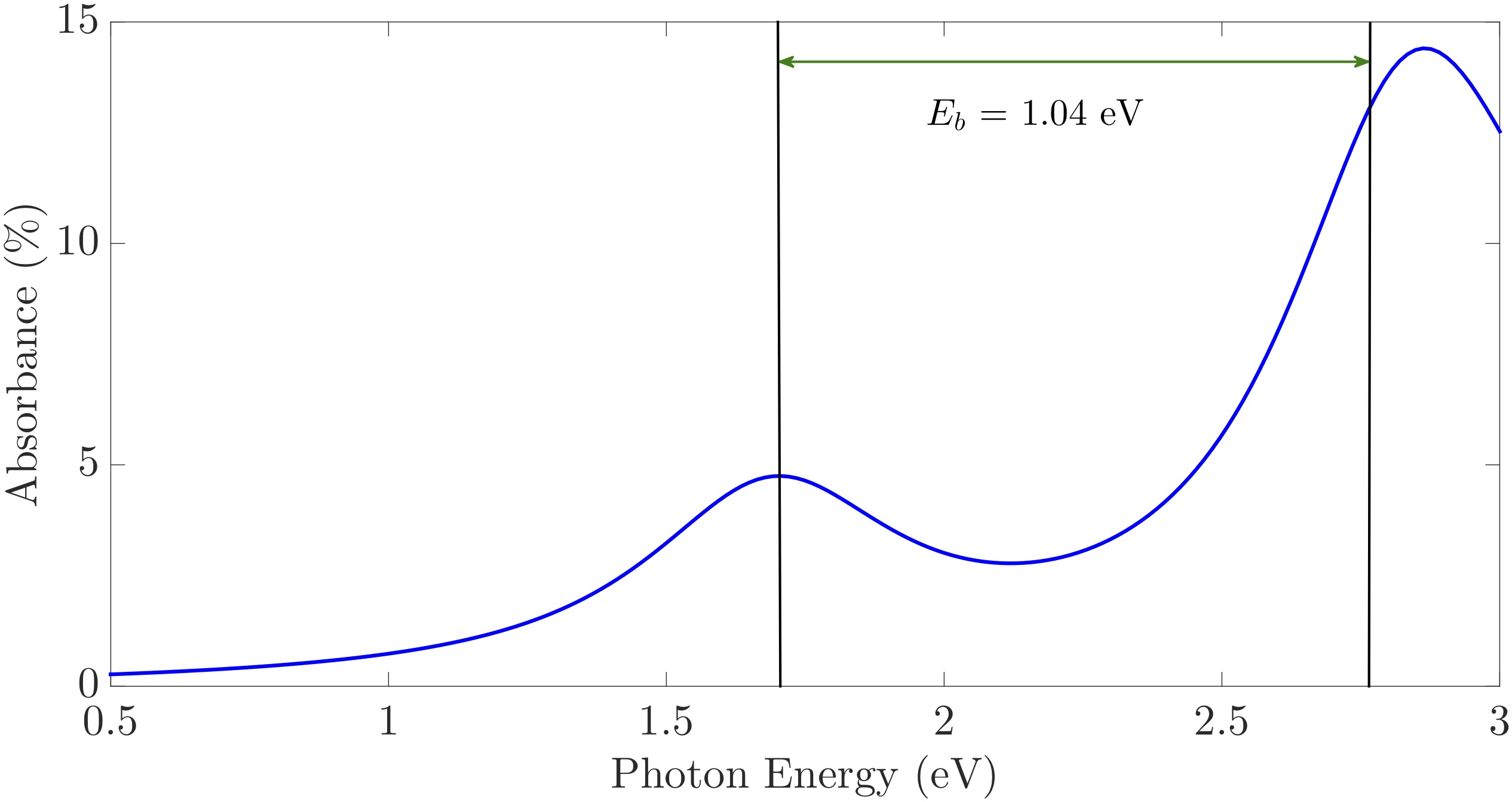}
\caption{Plot of the absorbance  (Eq. \ref{eq17}) using parameters for a $\text{MoS}_2$ TMD on a fused silica substrate:  energy gap $\Delta=2.82 \,$eV and background refractive index $n_b=1.5$.  We use a $\delta = 0.1\Delta$ Lorentzian broadening on both exciton and continuum transitions. $E_{b}$ is the binding energy for the $1s$ Dirac exciton.}
\label{fig1}
\end{center}
\end{figure}

The starting point for our analysis is the low-energy Hamiltonian for the band structure in the vicinity of the Dirac points ($\k=0$). Governed by the symmetry properties of the hexagonal lattice, the lowest order $\k \cdot \p$ Hamiltonian including the light-matter interaction and neglecting the spin-orbit interaction has the form 
\beq \label{eq1}
H_{\k} = v_F\Bigl[\boldsymbol{\sigma} \cdot\Bigl( \hat \p + \frac{e}c  \A(t)\Bigl)\Bigl] + \sigma_{z}\frac{\Delta}{2} +  I_2 V(\bold r),
\eeq
where $v_F$ is the Fermi velocity, $\A(t)$ is the vector potential, $\Delta$ is the energy gap, $I_2$ is the $2\times2$ identity matrix, $V(\bold r)$ is the electrostatic potential and $\boldsymbol \sigma = (\sigma_x,\sigma_y)\text{,} \, \sigma_z$ are the Pauli matrices. The usual approach used to study the optical properties of this Hamiltonian  parallels that of  the Semiconductor Bloch equations (SBEs) \cite{lk}. However  recent works  show that a full understanding of the nonlinear optics of Dirac materials requires one to go beyond the SBEs \cite{I,cm,cb}. In 2010 Ishikawa derived an extended version of the SBEs for  graphene using the formalism of  instantaneous eigenstates \cite{I} which has been recently generalised for a gapped material \cite{cb}.
The theory developed in these works unveil novel and previously unexplained nonlinear optical properties of graphene and gapped graphene, however they do not include Coulomb interactions that are in general very strong in two-dimensional semiconductors, preventing meaningful predictions to be made. The inclusion of Coulomb  interactions within this new formalism requires the second quantisation of Hamiltonian Eq. (\ref{eq1}).  To accomplish this, we need to quantise the instantaneous Dirac field; the procedure is simple but rather lengthy (see Supplementary Material) and it results in the second quantised Hamiltonian 
\beq \label{eq2ndQuant}
H = \sum_{\k}\epsilon_{\k}(t) (\ad\a - \bd\b) - \hbar \Omega_{\k}(t) (\ad\b + \text{h.c.}),
\eeq
where $a_{\k}$ and $b_{\k} $ are respectively the annihilation operators for electrons and holes, while $\epsilon_{\k}(t)=\pm\sqrt{v_F^2|(\p+e/c\A(t))|^2+\Delta^2/4}$ and $\Omega_{\k}(t)=ev_FE(t)/\hbar[ \sin \theta_\k/(2\epsilon_{\k}) - i \Delta \cos \theta_\k/(4\epsilon^2_{\k})]$ are the instantaneous energy and Rabi frequency respectively. This Hamiltonian differs from the usual second quantised  2D semiconductor case because both  $\epsilon_{\k}(t)$  and $\Omega_{\k}(t)$ explicitly depend on the interaction field $\A(t)$.  
The presence of the Coulomb potential $V(\r)$, leads to a renormalisation of  both the instantaneous energy and the Rabi frequency. Introducing the microscopic polarisation $p_{\k} = \langle \b\a \rangle$ and the electron-hole occupation numbers $n_{\k}^{e}= \langle \ad\a \rangle$, $n_{\k}^{h}= \langle \bd\b \rangle$ we can  show that they satisfy the following system of {\em renormalised Dirac-Bloch equations} (RDBs):
\begin{eqnarray} \label{eq2}
\hbar \dot p_{\k} &=& -2i~\epsilon_{\k}^R(t) p_{\k} - i\hbar\Omega^{R}_{\k}(t) e^{i2 \gamma_{\k}} (n_{\k}^{e}-n_{\k}^{h}), \nonumber \\
\dot n^{e,h}_{\k} &=& - 2~\text{Re}(\Omega^{R}_{\k}(t))~\text{Im}(p_{\k}e^{i2 \gamma_{\k}}) \nonumber \\
&& ~~~~~~~~~~~~~~~~~- 4 ~\text{Im}(\Omega^{R}_{\k}(t)) ~\text{Re}(p_{\k}e^{i2 \gamma_{\k}}),
\end{eqnarray}
where $\epsilon^{R}_{\k}(t) = \epsilon_\k(t) + \sum_{\k'\neq\k}V_{|\k-\k'|}(n_{\k'}^{e} - n_{\k'}^{h})$ and $\Omega^{R}_{\k}(t)=\Omega_{\k}(t) + 1/\hbar  \sum_{\k'\neq\k}V_{|\k-\k'|}p_{\k'}$ are the renormalised energy and Rabi frequency. The phase term ($\gamma_{\k}$) is a gap induced Berry phase (see Supplementary information for details). The formulation of Eqns \ref{eq2ndQuant}  and \ref{eq2} are the first results of this Letter. As was the case for the SBEs  \cite{lk,lb,da,zs}, the RDB  equations framework  can be flexibly applied to many  optical pulsed experiments and provide  a route to the microscopic understanding of  linear and nonlinear properties of gapped TMDs. This approach will pave the way for a complete understanding of the excitonic structure and the interaction of ultrashort, intense light pulses with Dirac materials, an area that at present is completely unexplored and potentially full of groundbreaking discoveries.

As a first example of the practical application of our new equations, we now consider the RDBs  in the low intensity and low density limit ($e/cA(t)<<\Delta/(2v_F)$ and  $n_{\k}^{e}-n_{\k}^{h}=-1$) where only the polarization equation survives. We will also assume a continuous wave (CW) radiation field. This will allow us to explore analytically excitonic effects on the linear optical properties of gapped TMDs.   Eq. (\ref{eq2}) then reduces to
\beq \label{eq6}
i\hbar \frac{d}{dt} p_{\k} = - 2 (\epsilon_\k + \sum_{\k'\neq\k}V_{|\k-\k'|}) p_{\k} + \hbar \Omega^{R}_{\k}(t) e^{i2 \gamma_{\k}},
\eeq 
with the non-renormalised energy  reducing to $\epsilon_\k = \sqrt{v_F^2\p^2+\Delta^2/4} $. To this equation is associated the well known Wannier stationary eigenvalue problem, that, for a massive Dirac quasiparticle reads 
\beq \label{eqWannier}
\Bigl[-iv_F \hbar \,  \boldsymbol{\sigma} \cdot \nabla + \sigma_z \frac{\Delta}2 + V(\bold r)\Bigl] \vec \Psi_{\nu}(\bold r) = E_\nu \vec \Psi_{\nu}(\bold r),
\eeq
here $\vec \Psi_{\nu}(\bold r)$ ($\nu = n,j)$ is a two component spinor eigenfunction of the electron-hole pair states which in polar coordinates has the form 
\beq \label{eqSpinor}
 \vec \Psi_{\nu}(\bold r) = \left(\begin{array}{c} ~~~e^{i(j+1/2)\phi_\r} ~F_{\nu}(r) \\  \pm i e^{i(j-1/2)\phi_\r} ~G_{\nu}(r) \end{array}\right),
\eeq
where $j=m+ 1/2$ is the eigenvalue of the "isospin-angular" momentum $\hat J_z = \hat L_z + \frac{1}2 \sigma_z$ along the $z$ axis and $n$ is the principal quantum number. The corresponding eigenvalues are 
\beq \label{eq8}
E_{\nu} = \hbar \omega_\nu = \frac{\Delta}{\sqrt{1+ \frac{\alpha_c^2}{(n+\gamma)^2}}},
\eeq
with $\gamma = \sqrt{j^2-\alpha^2_c}$. The constant $\alpha_c$ is the dimensionless Coulomb coupling strength and is determined by the background dielectric constant \cite{Novikov}. 

In what follows we use  the Wannier states in Eq. (\ref{eqSpinor})  to derive the electric susceptibility  and hence the linear absorbance; thereby recasting  the Elliott formula for gapped Dirac materials. To solve Eq. (\ref{eq6}) we  expand the microscopic polarisation in term of these Wannier states, after a lengthy but straightforward  calculation \cite{hk} we obtain, 
\begin{equation} \label{eq11}
\vec P(\bold r,\omega) = - \mathcal{E}(\omega) L^2 \sum_{\nu} \frac{\int  {\vec \mu_{cv}}(\bold r)\cdot\vec\Psi^{\dag}_{\nu}(\bold r) d\bold r}{\hbar \omega - \hbar\omega_{\nu} + i\delta} \vec \Psi_{\nu}(\bold r).
\end{equation}
Here $\vec \mu_{cv}(\bold r) = \mu_{cv}(\bold r) \left(\begin{array}{c}1 \\1\end{array}\right)$ is the electric dipole in spinor form, $\mathcal E(\omega)$  is the electric field in frequency domain, $L^2$ the area of the sample and $\delta$ is an energy broadening. We can now define the macroscopic polarization as follows 
\beq
P(\omega) = \int  [\vec \mu_{cv}^\dag(\bold r)\! \cdot\!\vec P(\bold r,\omega) + \vec \mu_{cv}(\bold r)\!\cdot\!\vec P^{\,\dag}(\bold r,-\omega)] \,d\bold r ,
\eeq
and inserting Eq. (\ref{eq11}) we get
\beq
P(\omega) = -L^2 \, \mathcal E(\omega) \sum_{\nu} \frac{ |\mathcal I_{\nu}|^2}{\hbar \omega - \hbar \omega_{\nu} +i\delta} ,
\eeq
where the oscillator strength is given by 
\beq \label{eq15}
|\mathcal I_{\nu}|^2 = \left| \int  \mu_{cv}(\bold r)(e^{i(j+1/2)\phi_{\r}} F_{\nu}(r)+ e^{i(j - 1/2)\phi_{\r}} G_{\nu}(r)) d \bold r \right|^2 .
\eeq
From the definition of  polarisation in c.g.s units 
$
P(\omega) =  L^2 d \,\chi(\omega)  \mathcal E(\omega),
$
we finally obtain the electron-hole pair susceptibility as 
\beq \label {eqchi}
\chi (\omega) = -\frac{1}{d } \sum_{\nu} \frac{|\mathcal I_{\nu}|^2}{\hbar\omega  - \hbar\omega_{\nu} + i\delta }.
\eeq
The nominal layer thickness $d$ has to be introduced to compute the volume within which the dipoles are induced but as we shall see it falls out in the calculation of the absorbance $\cal A$ which is the absorption coefficient multiplied by the layer thickness $d$.  The Influence of this effective layer thickness on the excitonic ground-state has been studied in a recent work \cite{ms}.\\
From Eq. (\ref{eqchi}), using the Plemelj-Sokhosky theorem, we obtain the final expression of the linear absorbance: the {\em relativistic Elliott formula} for Dirac materials 
\begin{widetext}
\beq \label{eq17}
{\cal A}(\Omega)  = \pi \alpha \Omega \Biggl[  \sum_{\nu} |\mathcal I_{\nu}|^2 \delta \Biggl( \Omega - \frac{1}{\sqrt{1+ \frac{\alpha_c^2}{(n+\gamma)^2}}}\Biggl) + C(\Omega)\  \Theta(\Omega-1)\Biggl],
\eeq 
\end{widetext}
where  $\Omega = \hbar \omega / \Delta$ is the scaled photon energy and $\Theta$ is the Heaviside step function. The two terms of Eq. (\ref{eq17}) are the discrete and the continuum contributions of the absorbance spectrum. Despite its striking similarity to the non-relativistic case Eq.(\ref{eq17})  is not trivial to evaluate since the integral given in Eq. (\ref{eq15}) can not be solved in closed form and must be evaluated numerically with particular care (some detail about the structure of the dipole moment are given in the Supplementary material). 
The factor $C(\Omega)$ in the second term of Eq. (\ref{eq17}) is what we may call the {\em relativistic Sommerfeld factor}: 
\beq \label{eq19}
C(\Omega) =\sum_{\nu} \frac{\Omega}{\sqrt{\Omega^2-1}}~|\mathcal I_\nu(\Omega)|^2.
\eeq
Where the sum here extend over the continuum states. This quantifies how the presence of electron-hole interactions enhances the continuum  absorbance with respect to the free carrier limit \cite{hk}.
It is maximum at the energy gap ($\Omega =1$) with a value that depends on the coupling constant $\alpha_c$ and tends to unity as the photon energy increases. This behavior is closely related to the relativistic energy spectrum and to the structure of the electron-hole continuum wave-function in contrast with the non-relativistic case which has a value of 2 at the bandedge. 
\begin{figure}[h]
\begin{center}
\includegraphics[width=1 \columnwidth]{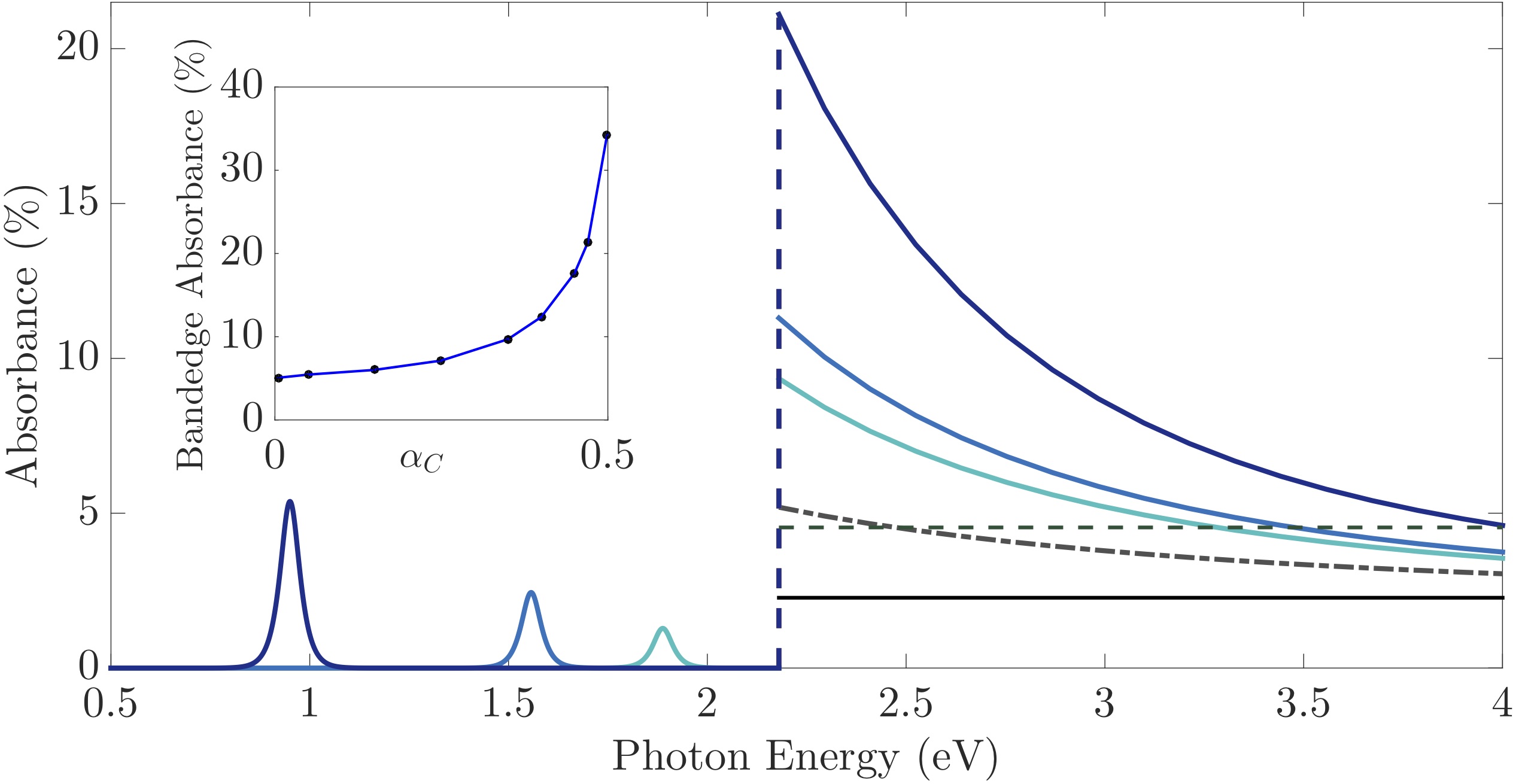}
\caption{Plot of the Elliott formula reducing the strength of the Coulomb coupling constant $\alpha_c$. Dark blue ($\alpha_c = 0.45$). Blue ($\alpha_c = 0.30$). Cyan ($\alpha_c = 0.20$). Free carrier limit (dashed-dotted grey line), zero gap graphene limit (dotted black line), non-relativistic limit (dashed green line). Inset: Continuum absorbance at the bandedge  ($\Delta=2.18 \,$eV) when increasing the strength of the Coulomb interactions.} \label{fig2}
\end{center}
\end{figure}

We now use Eq. (\ref{eq17}) to demonstrate the development of the Coulomb effects on the absorbance spectrum for varying dielectric screenings and energy gaps. 
In figure \ref{fig2} we plot the absorbance spectrum  while varying the strength of the Coulomb interaction $\alpha_c$. Experimentally it is possible to tune the strength of the Coulomb interaction by changing the substrate on which the sample is deposited since it changes the overall dielectric constant.  As expected the bound-state exciton lines  reduce their binding energy and oscillator strength as the Coulomb interaction weakens (i.e.  reducing $\alpha_c$).  For the continuum states, the Coulomb enhancement (Sommerfeld factor) is strongest at the bandedge, see Eq. (\ref{eq19}). Approaching the zero Coulomb limit ($\alpha_c=0$)  Eq. (\ref{eq17}) reproduces exactly the free carrier absorbance spectrum that can be computed analytically as  ${\cal A}(\Omega) = \pi \alpha \, (\Omega^2 + 1) / \Omega^2$.  
As can be seen from the dashed-dotted  line in figure \ref{fig2} this non-interacting model  is unable to  reproduce the giant  continuum absorbance seen in experiment. It is therefore clear that the giant absorption of Dirac media such as TMDs is due to Coulomb interactions.
The absorbance reduces quite rapidly as the photon energy is increased and at high photon energy the absorbance approaches the universal (graphene) absorbance value of 2.3\%. In this limit the high energy electron and hole scattering wavefunctions involved are essentially identical to the free particle wavefunctions. 
The inset in figure \ref{fig2} shows how the absorbance at the band-gap varies with $\alpha_c$. We can clearly observe that it increases dramatically as the coupling constant approaches the critical value ($\alpha_c = 0.5)$ where a phase transition to an excitonic insulator occurs \cite{sk,gs,rc}.  This phase transition is a fundamental signature of purely Dirac excitons and it potentially represents a measure of the \emph{Diracness} \cite{gm} of the two body electron-hole system since it disappears for a parabolic exciton dispersion \cite{tg}.  
\begin{figure}[h]
\begin{center}
\includegraphics[width=1 \columnwidth]{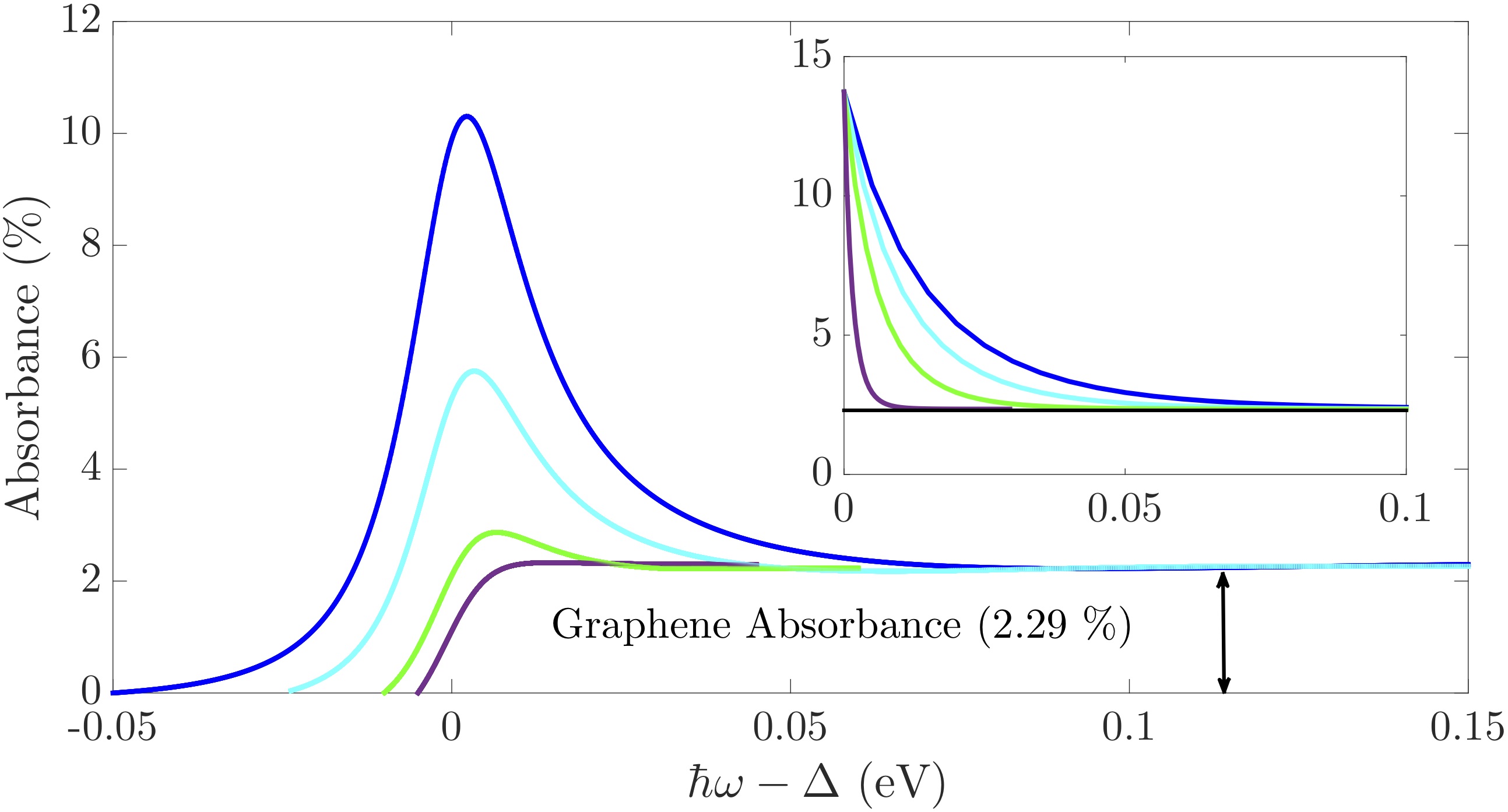}
\caption{Plot of continuum absorbance contribution while reducing the bandgap $\Delta$,  ($\alpha_c=0.35$) with a finite Lorentzian energy broadening $\delta = 10\,$meV. Blue: $\Delta = 100\,$ meV. Cyan: $\Delta = 50\,$ meV. Green: $\Delta = 25\,$ meV. Violet: $\Delta = 5\,$ meV. Inset: continuum absorbance  absorbance  while reducing the bandgap for zero energy broadening ($\alpha_c=0.35$). The axes labels are the same as in the main plot.} \label{fig4}
\end{center}
\end{figure}

One of the most remarkable features of graphene from a physics point of view is that there appears to be no Coulomb enhancement of the continuum absorbance, despite the existence of final state Coulomb interactions. Figure \ref{fig4} shows the continuum absorbance calculated from  Eq. (\ref{eq17}) varying the energy gap ($\Delta$). We  observe that reducing the bandgap the curves approach  the universal absorbance value of graphene $\pi \alpha$ at lower and lower photon energies.  This indicates that the relevance of Coulomb interactions is confined to a region close to the energy gap and once the gap is closed the system behaves as a nearly free electron gas as seen in experiments \cite{nb} and in the Das Sarma paper \cite{hd} by using a renormalisation group approach. Within our theoretical approach this phenomenon can be explained by observing that in the relativistic hydrogenic model the energy gap is the exact analog of the Rydberg constant  that is the only relevant energy scale.  This means that tuning the gap changes the relevant energy scale of the system; when it is reduced to zero all photon mediated transitions can be considered as high energy transitions and in this regime the system it is known to reproduce the free absorbance limit. 

In conclusion we studied the linear optical properties of two-dimensional gapped Dirac materials. We derived a new system of Coulomb-renormalised Dirac Bloch equations based on the quantisation of instantaneous eigenstates. We provided an explicit expression of the absorbance Elliott formula that  to the best of our knowledge was still missing in previous theoretical studies.  We demonstrated  how the absorbance spectrum evolves as the energy gap and the strength of Coulomb interaction are reduced to zero. In particular we showed that even in the presence of relatively strong electron-hole Coulomb interactions the gap reduction leads to universal absorbance and we relate this to the structure of the relativistic hydrogenic problem. Our new theoretical framework will be potentially used to predict the behaviour of the interaction of ultrashort pulses with Dirac media in a variety of experimental situations, leading to a fresh understanding on new mechanisms of higher-harmonics generation and the study of cavities, multilayers and TMD-enhanced microresonators.

LDMV acknowledges support from EPSRC under the auspices of the Scottish Centre for Doctoral Training in Condensed Matter Physics.

\newpage
\begin{widetext}
\section{Supplementary information}

%
%
%
%
%
\def \hP{\hat \psi}
\def \hPd{\hat \psi^\dag}
\def \hPb{\hat \bar \psi}
\def \slPar{\slashed \partial}
\def \D{\slashed D}
\def \beq{\begin{equation}}
\def \eeq{\end{equation}}
\def \bes{\begin{split}}
\def \ees{\end{split}}
\def \besu{\begin{subequations}}
\def \esu{\end{subequations}}
\def \beal{\begin{aligned}}
\def \eal{\end{aligned}}
\def \ac{\alpha_c}
\def \g{\gamma}
\def \G{\Gamma}
\def \vsigma{\underline \sigma_T}  
\def \vpi{\underline \pi}
\def \vP{\vec \psi}
\def \r{\bold r}
\def \x {\bold x}
\def \y{\bold y}
\def \k{\bold k}
\def \p{\bold p}
\def \A{\bold A}
\def \a{\hat a_{\k}}
\def \ad{\hat a^{\dagger}_{\k}}
\def \b{\hat b_{-\k}}
\def \bd{\hat b^{\dagger}_{-\k}}
\def \ap{\hat a_{\k'}}
\def \adp{\hat a^{\dagger}_{\k'}}
\def \bp{\hat b_{-\k'}}
\def \bdp{\hat b^{\dagger}_{-\k'}}
\def \bea{\begin{eqnarray}}
\def \eea{\end{eqnarray}}

\title{Coulomb effects in the absorption spectra of two dimensional Dirac materials: Supplementary informations}
\author{Leone Di Mauro Villari}
\author{Ian Galbraith}
\author{Fabio Biancalana}
\affiliation{Institute of Photonics and Quantum Sciences, School of Engineering and Physical Sciences, SUPA, Heriot-Watt University, Edinburgh EH14 4AS, UK.}

\maketitle 

\section{Quantisation of the Instantaneous Dirac Field and Renormalised Dirac-Bloch Equations}
The starting point of our analysis is the low-energy Hamiltonian for the band structure in the vicinity of the Dirac points. Dictated by the symmetry properties of the hexagonal lattice, the lowest order $\k \cdot \p$ Hamiltonian including the light-matter interaction has the form
\beq \label{eq1}
H^{\xi}_\k = H^\xi_{\k,D}+ H_C = v_F\Bigl[\boldsymbol{\sigma}_{\xi} \Bigl( \p + \frac{e}c  \A(t)\Bigl)\Bigl] + \sigma_z\frac{\Delta}{2}  + \hat I_2 V(\r)
\eeq
where $A_{\mu}(\bold x,t) = (\A(t),V(\r))$ is the three-vector potential, $\xi$ is a valley index referring to the Dirac points ($\bold K$, $\bold K'$) and $\boldsymbol \sigma = (\xi \sigma_x,\sigma_y)\text{,} \, \sigma_z$ are the Pauli matrices. Since we are not considering the effect of the spin-orbit coupling we can restrict to a single valley, namely $\xi=1$. The goal of this section is to derive a renormalised system of Dirac-Bloch (RDB) equations from the second quantised form of the Hamiltonian in eq. (\ref{eq1}). In order to do so we consider the following interacting Dirac equation
\beq
i\hbar \Pdert \psi(\r,t) = \Bigl\{v_F\Bigl[\boldsymbol {\sigma} \Bigl( i\hbar\mathbf \nabla + \frac{e}c  \A(t)\Bigl)\Bigl] + \sigma_3\frac{\Delta}{2}\Bigl\} \psi(\r,t).
\eeq
 An ansatz solution to this equation can be given in terms of instantaneous eigenstates of the secular equation 
 \beq
 H_{\k,D} u_{\lambda,\k}(t) = \epsilon_{\lambda,\k} u_{\lambda,\k}(t).
 \eeq
 These eigenstates can be written in the normalised from
\beq \label{spinor}
  \vec u_{\lambda,\k}(t) = \frac{v_F|\bold \pi_k|}{\sqrt{\epsilon_{\k}(\lambda \Delta + 2\epsilon_\k)}}\left(\begin{array}{c}\Biggl(\frac{\lambda\Delta + 2\epsilon_{\k}}{2v_F|\pi_{\k}|}\Biggr) e^{-i\phi_{\k}/2} \\e^{i\phi_{\k}/2}\end{array}\right),
 \eeq
 here $|\boldsymbol \pi_\k|=\p + \frac{e}c \A(t)$, $\theta_\k = \arctan(\pi_x/\pi_y)$ and $\lambda=\pm 1$ labels the conduction and valence bands. $\epsilon_\k(t)$ is  the positive branch of the instantaneous energy. 
\beq
\epsilon_{\lambda,\k} = \lambda \sqrt{\Biggl(\frac{\Delta}2\Biggl)^2 + (v_F| \boldsymbol \pi_\k|)^2}.
\eeq
The addition of the gap leads to an inequivalence of $\bold K$ and $\bold K'$ sublattices and consequently to the appearance of a Berry phase $\gamma_{\lambda,\k}(t) = \int_{-\infty}^t  d\tau \,  \vec {\dot u}^{\,\dagger}_{\lambda,\k}(\tau) \! \cdot \! \vec u_{\lambda,\k}(\tau) $. Consequently the spinor (\ref{spinor}) evolves in time as follows
\beq
\vec \psi_{\lambda,\k}(\r,t) = \vec{u}_{\lambda,\k}(t) e^{-\lambda i(\theta_{\k}(t) - \gamma_\k(t))+i\k\r}.
\eeq
 Where $\theta_\k(t) = \frac{1}\hbar \int_{-\infty}^t \epsilon_\k(\tau)d\tau$ is the \emph{dynamical} phase and $\gamma_\k(t)$ the positive branch Berry phase. The Dirac field $\psi(\r,t)$ can be then expanded in terms of instantaneous eigenstates 
 \beq
 \vec \psi(\r,t) = \sum_{\k,\lambda=\pm 1} \, a_{\k,\lambda} \vec u_{\k,\lambda} e^{-\lambda i(\theta_{\k}(t) - \gamma_\k(t))+i\k\r},
 \eeq
with $a_{\k,\lambda}$ are band electron ladder operators. Substituting this field in the Hamiltonian
\beq
H_{D} = \int d\r \, \vec \psi^\dagger(\r,t) \Bigl\{v_F\Bigl[\boldsymbol{\sigma}_{\xi} \Bigl( \nabla + \frac{e}c  \A(t)\Bigl)\Bigl] + \sigma_3\frac{\Delta}{2}\Bigl\} \vec \psi(\r,t),
\eeq 
we get
\beq
H_D = \frac{\hbar}{L^2}\int d\r \sum_{\k\k',\lambda} \hat a^\dagger_{\k',\lambda} u^\dagger_{\k',\lambda} e^{-i\k'\r + i\lambda \eta(t)}[a_{\k,\lambda}(i\dot u_{\k,\lambda} + u_{\k,\lambda} \dot\eta(t))e^{i\k\r -i\lambda \eta(t)} ] ,
\eeq
with $\eta(t) = \theta_\k(t) + \gamma_\k(t)$ being the sum of the Berry phase and the dynamical phase. Introducing the hole operators $b^\dagger_{-\k} = a_{\k,-1}$, after a lengthy but simple algebraic calculation we get the final form of the second quantised Hamiltonian
\beq \label{eq10}
H_D = \sum _\k [\epsilon_\k(t) \omega_k(t) (\ad \a - \b\bd)  - \hbar \Omega_\k (t) (\ad \b + \bd\a)]
\eeq
where $\Omega_\k(t)$ is a generalised  Rabi frequency and it is given by
\beq
\Omega_\k(t) = -i \vec u_{\k,1}^\dagger \cdot \vec{\dot u}_{\k,-1} = \frac{ev_FE(t)}\hbar\Bigl[ \frac{\sin \theta_\k}{2\epsilon_{\k}} - i \Delta \frac{\cos \theta_\k}{4\epsilon^2_{\k}}\Bigl].
\eeq
Introducing the Coulomb interactions in the Hamiltonian we get
\beq
\begin{split}
H =&  \sum _\k [\epsilon_k(t) (\ad \a - \bd\b)  - \hbar \Omega_\k (t) (\ad \b + \bd\a)] + \\ & \frac{1}{2} \sum_{\k,\k'}\sum_{\p \neq 0} V_\p (\hat a^\dagger_{\k +\p} \hat a^\dagger_{\k'-\p}\ap\a + \hat b^\dagger_{\k'+\p}\hat b^\dagger_{\k'-\p}\bp\b - 2\hat a^\dagger_{\k +\p} \hat b^\dagger_{\k'-\p}\bp\a),
\end{split}
\eeq
we can now derive the equations for the population and inversion variables $n^e_{\k} = \braket{\ad\a}$, $n^h_{\k}=\braket{\bd\b}$ and $p_\k = \braket{b_{-\k} a_\k}$. This lengthy procedure is quite standard and can be found in any book of semiconductor theory; for example \cite{hk1}. It results in the following set of RDB equations  
\beq \label{eq2}
\begin{aligned}
& \hbar \dot p_{\k} = -2i\epsilon^R_{\k}(t) p_{\k} - i \hbar \Omega_{R}(t) e^{i2 \gamma_{\k}} (n_{\k}^{e}-n_{\k}^{h}), \\
&\dot n^{e,h}_{\k} = - 2\text{Re}(\Omega_{R}(t))\text{Im}(p_{\k}e^{i2 \gamma_{\k}}) - 4 \text{Im}(\Omega_{R}(t)) \text{Re}(p_{\k}e^{i2 \gamma_{\k}}),
\end{aligned}
\eeq
where $\epsilon^R_{\k}(t) = \epsilon_\k(t) + \sum_{\k'\neq\k}V_{|\k-\k'|}(n_{\k'}^{e} - n_{\k'}^{h})$ and $\Omega_{R}(t)=\Omega_{\k}(t) + 1/\hbar  \sum_{\k'\neq\k}V_{|\k-\k'|}p_{\k'}$ are the renormalised energy and Rabi frequency. 

\section{The Wannier-Dirac equation}
 
 As in the case of 2D semiconductors the Wannier equation can be derived from the RDB equations (\ref{eq2}). We consider the system under the approximation $e/cA(t) << \epsilon_g/(2v_F)$  and under the low excitation approximation $w_k \approx -1$ Thus Eq. (\ref{eq2}) simplifies to 
\beq \label{eq3}
i \hbar \dot p_{\k}  =  -(2\epsilon_{\k}  + \sum_{\k'\neq \k} V_{|\k-\k'|}) p_{\k} + \hbar \Omega_\k(t) e^{2i\gamma_\k} , 
\eeq
with the non-renormalised energy reducing to $\sqrt{v_F \bold p^2 + \Delta^2/4}$. This equation represents a two dimensional relativistic two particle system with an inhomogeneous term given by the optical field. In order to define the differential operator in the square root we write Eq. (\ref{eq3}) as follows 
 \beq \label{3}
 i \hbar \frac{\partial}{\partial t} q_{a}(\k,t)  = [\hbar v_F (\alpha_i)_{ab} k_i + \beta_{\mu\nu}M v_F^2 ] q_{b}(\k,t) + i\sum_{\k'\neq \k}V_{|\k'-\k|}q_{a}(\k',t) - E(t)\mu_{\k} \gamma_{a}   
 \eeq
 where $\hat \alpha_i$, $\hat \beta$ are matrices, $\vec \gamma$ a spinor and the mass parameter is $M = \Delta/(2v^2_F)$. The dispersion in Eq. (\ref{3}) must satisfy
\beq 
\hbar v_F (\alpha_i)_{ab} k_i + \beta_{ab}M v_F^2  = \epsilon_{\k}
\eeq
Thus we must have 
\beq 
\{\hat \alpha_i,\hat \alpha_j\} = \delta_{ij}  \quad  (\hat \alpha_i)^2 = (\hat \beta)^2 = \hat I_2
\eeq
This algebra is satisfied by taking $(\tilde \alpha_1,\tilde \alpha_2) = (\sigma_x,\sigma_y) = \vsigma$ and $\tilde \beta = \sigma_z$. To determine the spinor $\vec \gamma$ we can switch off the Coulomb potential and impose that each component $q_\mu$ satisfy a Klein-Gordon equation with an inhomogeneous term $\mu_k E(t)$.  To accomplish this it is sufficient to take $\vec \gamma = \left(\begin{array}{c}1 \\1\end{array}\right)$.\\
Taking the Fourier transform of equation Eq (\ref{3}) we get 
\beq \label{4}
 i \hbar \frac{\partial}{\partial t} \vec q(\r,t)  = [-iv_F \hbar  \boldsymbol \sigma  \nabla + \sigma_z M v_F^2 + V(r)] \vec q(\r,t)  + E(t)\vec \gamma \mu(\r) 
\eeq 
We shall now solve the homogeneous stationary eigenvalue problem 
\beq \label{5}
[-iv_F \hbar  \boldsymbol \sigma  \nabla + \sigma_z M v_F^2 + V(r)] \vec \Psi_\nu(\r) = E_n \vec \Psi_\nu(\r)
\eeq
Where $\nu$ here indicates a set of  quantum numbers. The solution to this equation is known (see for example \cite{Novikov}) and we report it here for completeness. It is convenient to solve the problem in polar coordinate.  Writing the spinor in the form 
\beq \label{6}
\vec \Psi_\nu (\r)=  \left(\begin{array}{c}\varphi_\nu \\ \chi_\nu\end{array}\right)
\eeq
 We can write Eq. (\ref{5}) in spinor components (we will drop for the moment the quantum numbers subscript)
 \beq
 \begin{aligned}
 (E_c - k_c - U(r)) \phi &= (\partial_x - i\partial_y) \chi \\
 (E_c + k_c - U(r)) \chi &=  (\partial_x + i\partial_y) \phi
 \end{aligned} 
 \eeq
where we introduced the "Compton" energy $E_c = E_n/(\hbar v_F) $, the Compton wave vector $k_c = Mv_F/\hbar$ and $U = V/(\hbar v_F) = - e/(\hbar v_F r)= - \alpha_c / r$. The coupling constant $\alpha_c$  is computed under the random phase approximation i.e $\alpha_c = \frac {\alpha_0}{\varepsilon_{\text{RPA}}}$ with $\alpha_0 = e^2/(v_F\hbar)$. \\ In polar coordinates we have $i\partial_x \pm i\partial_y = e^{i\phi_{\r}}(\partial_r \pm 1/rL_z)$ where $L_z$ is the third component of the angular momentum operator $\underline L =(0,0,L_z)$. We can write now the spinor in Eq. (\ref{6})  in the form 
\beq \label{7}
\vec \Psi (\r) = \left(\begin{array}{c} e^{i(j+1/2)\phi_\r} F(r) \\  \pm i e^{i(j-1/2)\phi_\r} G(r) \end{array}\right)
\eeq
Where $j=m+ 1/2$ is the eigenvalue of the "isospin-angular" momentum $\hat J_z = \hat L_z + \frac{1}2 \sigma_z$ along the $z$ axis. Substituting the spinor (\ref{7}) in Eq.  (\ref{5}) we get the radial equation
\beq \label{8}
\begin{aligned}
\frac{d}{dr} F - \frac{j}{r} F + (E_c + k_c - U)G&=0,\\
\frac{d}{dr} G - \frac{j}{r} G + (E_c - k_c - U)F&=0.
\end{aligned}
\eeq
In spite of its apparent simplicity the solution of this system is far from being trivial. One way to solve it is to decouple it in two independent second order differential equations, to do so we write the solution in the following form 
\beq \label{9}
\begin{aligned}
F(r) &= (k_c + E_c)^{1/2} e^{-\lambda r} (2\lambda r)^{\gamma - 1/2} \tilde F(r)\\
G(r) &= (k_c - E_c)^{-1/2} e^{-\lambda r} (2\lambda r)^{\gamma + 1/2} \tilde G(r)
\end{aligned}
\eeq
with $\kappa = \sqrt{k_c^2 - E_c^2}$ and $\gamma = \sqrt{j^2 - \alpha_c^2}$.  Introducing the dimensionless radius  $\rho = 2 \lambda r$ we get 
\beq \label{10}
\begin{aligned}
\rho \frac{d}{d\rho} \tilde F + (\gamma -j)\tilde F - \frac{\rho}2 (\tilde F - \tilde G) + \frac{\kappa \alpha_c}{k_c + E_c} \tilde G &= 0, \\
\rho \frac{d}{d\rho} \tilde G + (\gamma + j)\tilde G - \frac{\rho}2 (\tilde F - \tilde G) + \frac{\kappa \alpha_c}{k_c - E_c} \tilde G &= 0.
\end{aligned}
\eeq 
Writing with $\tilde F = Q_1 + Q_2$ and $\tilde G = Q_1 - Q_2$  we find 
\beq \label{11}
\begin{aligned}
\rho \frac{d}{d\rho} Q_1 + (\gamma - \frac{\alpha_c E_c}\kappa)Q_1 - (j + \frac{k_c\alpha_c}\kappa)Q_2 &= 0, \\
\rho \frac{d}{d\rho}  Q_2 + (\gamma - \rho + \frac{\alpha_c E_c}\kappa)Q_2 - (j - \frac{k_c\alpha_c}\kappa)Q_1 &= 0,
\end{aligned}
\eeq 
looking at this system at $\rho = 0$ we derive the useful relation
\beq \label{12}
\gamma^2 - \Bigl(\frac{\alpha_c E_c}{\kappa}\Bigl)^2 = j^2 - \Bigl(\frac{k_c\alpha_c}{\kappa}\Bigl)^2.
\eeq
Using (\ref{12}) we can easily decouple the system, obtaining 
\beq \label{13}
\begin{aligned} 
\rho \frac{d^2}{d\rho^2} Q_1 + (1+2\gamma -\rho) \frac{d}{d\rho} Q_1 - \Bigl(\gamma - \frac{\alpha_c E_c}{\kappa} \Bigl) Q_1&=0, \\
\rho \frac{d^2}{d\rho^2} Q_2 + (1+2\gamma -\rho) \frac{d}{d\rho} Q_2 - \Bigl(1+\gamma - \frac{\alpha_c E_c}{\kappa} \Bigl) Q_2&=0.
\end{aligned}
\eeq
These equations are in the confluent hypergeometric form
\beq
z^2  \frac{d^2}{dz^2} f + (b-z) \frac{d}{dz} f - af=0.
\eeq
One solution is given by Kummer hypergeometric function the  ${}_1\mathcal{F}_1(a,b;z)$ \cite{olv}. Thus $Q_1$ and $Q_2$ are given by 
\beq \label{14}
\begin{aligned}
Q_1 &= C^{(1)} {}_1\mathcal F_1(\gamma - \alpha_c E_c/\kappa,1+2\gamma;\rho) \\
Q_2 &= C^{(2)} {}_1\mathcal F_1(1+ \gamma - \alpha_c E_c/\kappa,1+2\gamma;\rho)
\end{aligned}
\eeq
Since ${}_1\mathcal F_1(a,b,0) = 1$ looking again at  (\ref{11})  for $\rho = 0$ we get 
\beq
c^{(12)} = \frac{C^{(1)}}{C^{(2)}} = \frac{\gamma - \alpha_c E_c/\kappa}{j + k_c \alpha_c/\kappa}
\eeq
Thus we can write the eigenfunctions for the discrete spectrum ($\nu = n,j)$ 
\beq
\begin{aligned}
F_\nu & = (k_c + E_c)^{1/2} e^{-\rho/2} \rho^{\gamma - 1/2} C^{(1)}[{}_1\mathcal F_1(\gamma-  \alpha_c E_c/\kappa,1+2\gamma;\rho) + c^{(12)}{}_1\mathcal F_1(1+ \gamma - \alpha_c E_c/\kappa,1+2\gamma;\rho)] \\
G_\nu &  = (k_c - E_c)^{1/2} e^{-\rho/2} \rho^{\gamma - 1/2} C^{(1)}[{}_1\mathcal F_1(\gamma-  \alpha_c E_c/\kappa,1+2\gamma;\rho) -  c^{(12)}{}_1\mathcal F_1(1+ \gamma - \alpha_c E_c/\kappa,1+2\gamma;\rho)] \\
\end{aligned}
\eeq
The bound states occur when the first argument of the hypergeometric function is a negative integer i.e. 
\beq
\gamma-  \frac{\alpha_c E_c^\nu}\kappa = -n
\eeq
from this equation we directly  obtain the discrete spectrum 
\beq
E^\nu_c = \frac{k_c}{\sqrt{1+\frac{\alpha_c^2}{(n+\gamma)^2}}}
\eeq 
Using Mathematica we can find the normalisation factor from the condition 
\beq
 \int d\r \, \psi^\dagger(\r)\psi(\r) = 2\pi \int_0^\infty rdr\, (F^2+G^2) = 1
 \eeq
 we get 
 \beq
 C^{(1)} = \frac{(-1)^n \kappa^{3/2}}{2\pi k_c\Gamma(1+2\gamma)} \sqrt{\frac{\Gamma(1+2\gamma + n)(j+k_c\alpha_c/\lambda)}{\alpha_c n!}} 
 \eeq
We now compute the continuum states of the system by analytic continuation of the eigenfunctions of the Wannier equation in the region $|E|> \Delta/2$, i.e. for complex values of $\kappa$. Switching back to Compton units the analytic continuation in the $E_c$ complex plane is easily accomplished by
\beq
\begin{aligned}
\kappa = \sqrt{k_c^2-E_c^2} \to -ik, \quad k = \sqrt{E_c^2 - k_c^2} \\
c_{12} \to e^{-2i\xi_j} = \frac{\g - i\ac^E}{j+iM\ac/k}, \quad \ac^E = \frac{\ac E}{k} 
\end{aligned}
\eeq
Substituting this prescription in the eigenfunctions after lenghty but straightforward calculations \cite{Novikov1} we get 
\beq
\begin{aligned}
F &= 2\sqrt{\frac{|E_c+k_c|}{\pi E_c}} \frac{|\G(1+\g+i\ac^E|)}{\G(1+2\g)} e^{\pi\ac^E/2} (2 k r)^{\g-1/2} \text{\emph{Re}}\{e^{ikr+i\xi} {}_1\mathcal F_1(\g-i\ac^E, 1+2\g;-2ikr)\}\\
G &=\pm 2\sqrt{\frac{|E_c-k_c|}{\pi E_c}} \frac{|\G(1+\g+i\ac^E|)}{\G(1+2\g)} e^{\pi\ac^E/2} (2 k r)^{\g-1/2} \text{\emph{Im}}\{e^{ikr+i\xi} {}_1\mathcal F_1(\g-i\ac^E, 1+2\g;-2ikr)\}.
\end{aligned}
\eeq
Both the discrete and continuum eigenstates derived here are used in the paper for the calculation of the Elliot formula.
\section{The optical matrix element}
It is useful  to compute the dipole moment in the position space. Using the more manageable Compton units we get
\beq
\mu(\r) = \frac{L^2}{(2\pi)^2}\int d\k \, \mu_{\k} e^{i\k\,\r} =   e \frac{L^2}{2\pi^2}\int_0^{\infty} k dk\,  \int_0^{2\pi} d\phi_\r e^{ikr(\phi_\k-\phi_\r)}  \Biggl(\frac{ \sin \phi_\k }{\sqrt{ k^2 + k_c^2}} - i k_c\frac{\cos \phi_\k}{k^2+ k_c^2}\Biggl)
\eeq 
This integral converges and can be solved analytically. With Mathematica$^\copyright$ we get 
\beq \label{dip}
\mu(\r) =e  \frac{L^2}{(2\pi)^2}\Bigl[ k_c\pi^2 \cos (\phi_\r) [I_1(k_cr) - S_{-1}(k_cr)] -  i  \sqrt{\pi} k_c \sin(\phi_\r) G\Bigl(0,-\frac{1}2,\frac{1}2,\frac{1}2\Bigl |\frac{(k_cr)^2}{4} \Bigl)\Bigl]
\eeq
Where $G(a_n,b_n|z)$ is the Mejer G function and $S_\alpha(z)$ is the the Struve function of order $\alpha$. In the case of zero gap ($k_c \to 0$) the integral simplifies considerably giving
\beq
\mu(\r) =  i\frac{L^2}{(2\pi)^2} \frac {\sin \phi_\r} r
\eeq
The dipole moment in eq (\ref{dip}) is essential to compute  the oscillator strength $|\mathcal I_\nu|^2$ defined in the integral (11) of the main text. This integral is not trivial to compute due to the special functions in the dipole moment. In particular even if both the real and imaginary part of $\mu(\r)$ converge \cite{olv}, the numerical implementation is problematic in particular in the region $ r >>1$. A way to solve this problem is to find a fit for these functions, in the large $r$ region, involving just simple polynomials and exponentials. Using Mathematica$^\copyright$ we find the following best fit
\beq
\beal
&\frac{\mu_R(r)}{eL^2} \to a e^{-b k_cr} \quad &a = -0.048592 \,\, b = -0.780094, \\
&\frac{\mu_I(r)}{eL^2} \to c e^{-k_cr} + d\frac{e^{-k_cr}}{(k_cr)^2} + \frac{e}{(k_cr)^2} \quad &c = -0.021511 \,\, d = - 0.071833 \,\, e = 0.027752,
\eal
\eeq
where with $\mu_{R,I}(r)$ we indicate the radial components of the real and imaginary part of the dipole moment. To compute the oscillator strength we thus use the exact dipole moment between $0$ and $k_c r_{cut} = 2$ and the fit between $k_cr_{cut}=2$ and $ k_cr_{cut} = 30$. We verified that integrating up to $k_c r_{cut} = 30$ was enough for the numeric integral to converge.
\begin{figure}[H]
\begin{center}
\includegraphics[width=0.7 \columnwidth]{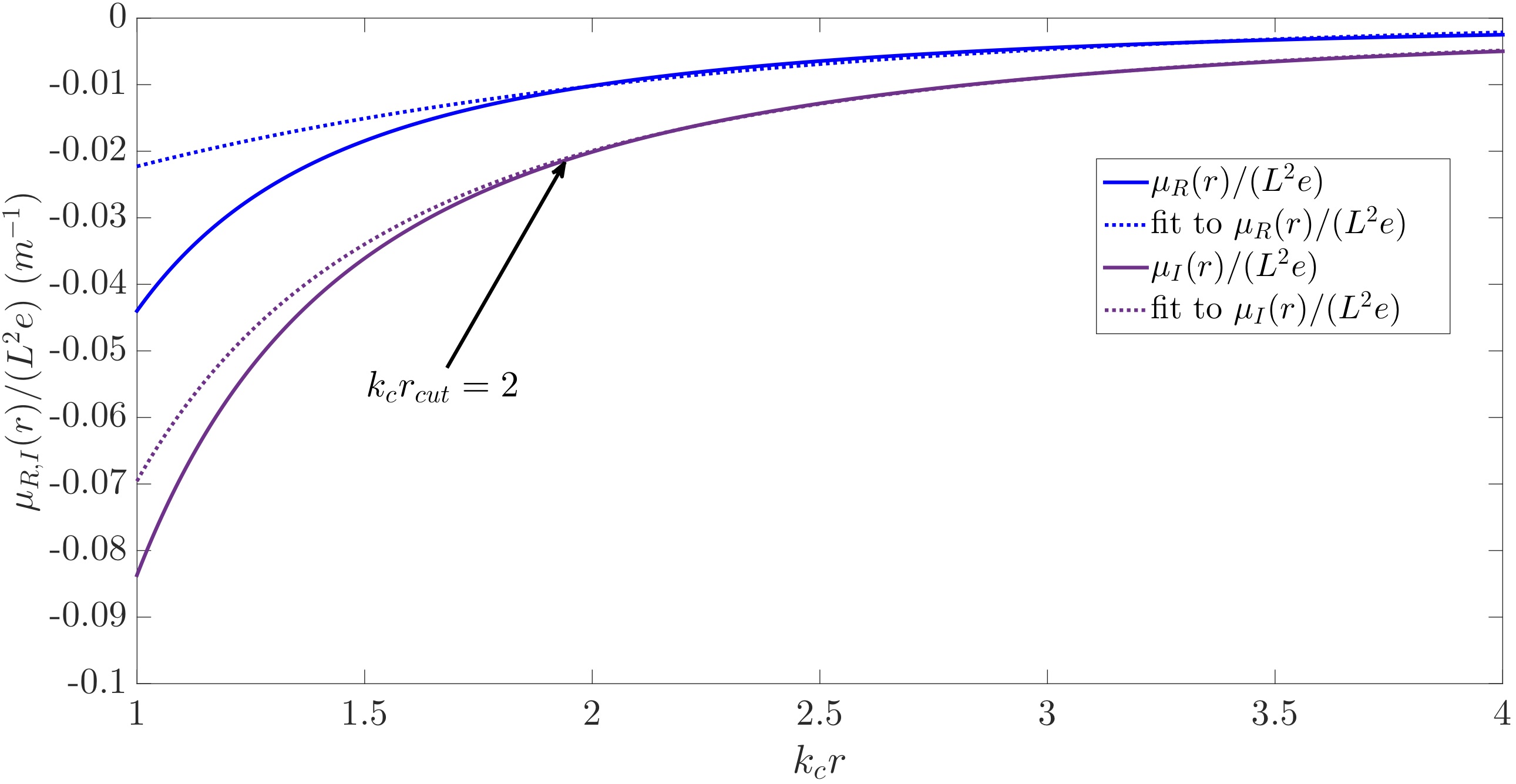}
\caption{Fit of the  the radial components of the real and imaginary part of the  dipole moment in units of area ($L^2$) and electric charge ($e$).}
\label{default}
\end{center}
\end{figure}

\section{Comparison with Experiments}
We shall now compare the results of our theoretical predictions with previous experimental results. To do so we compute the real and imaginary part of the dielectric function for MoSe$_2$ and MoS$_2$ on a fused silica substrate and we compare our calculations with the experimental results in \cite{lc}. The complex dielectric function is simply given by 
\beq
\epsilon(\omega) = \epsilon_b + \chi(\omega)
\eeq
where $\epsilon_b$ is the background dielectric constant of the substrate, here we assume a thickness $d = 1 \, nm$ for the monolayers.
\begin{figure}[H]
\begin{center}
\includegraphics[width=1 \columnwidth]{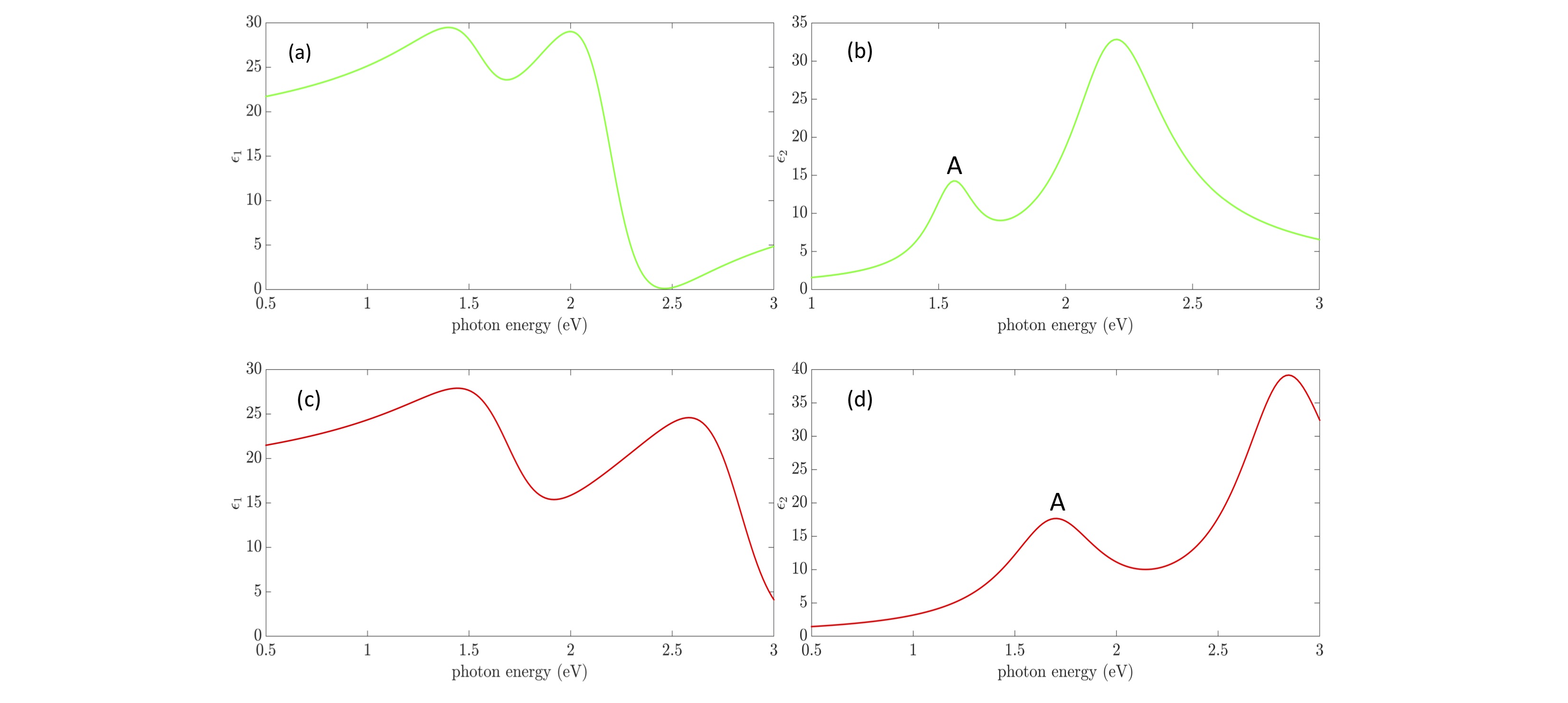}
\caption{(a) Real part of the dielectric function of MoSe$_2$, (b) Imaginary part of the dielectric function of MoSe$_2$, (c) Real part of the dielectric function of MoS$_2$, (d)  Imaginary part of the dielectric function of MoS$_2$. This figure show a good agreement with the results in figure 3\emph{a,i,c,k} of \cite{lc1} }
\label{default}
\end{center}
\end{figure}



\end{widetext}
\end{document}